%% file: main.tex
\def\year{2019}\relax
\documentclass[letterpaper]{article}
\usepackage{aaai}
\usepackage{times}
\usepackage{helvet}
\usepackage{courier}
\usepackage[hyphens]{url}  
\usepackage{graphicx} 
\usepackage{graphicx}  
\title{Privacy, altruism, and experience: \\ Estimating the perceived value of Internet data for medical uses} 

\author{
Gilie Gefen\\Technion\\gilie@campus.technion.ac.il
\And Omer Ben{-}Porat\\Technion\\omerbp@campus.technion.ac.il  
\AND Moshe Tennenholtz\\Technion\\moshet@ie.technion.ac.il
\And Elad Yom{-}Tov\\Microsoft Research Israel and Technion\\eladyt@microsoft.com
}	

\begin{document}
\maketitle

\begin{abstract}
People increasingly turn to the Internet when they have a medical condition. The data they create during this process is a valuable source for medical research and for future health services. However, utilizing these data could come at a cost to user privacy. Thus, it is important to balance the perceived value that users assign to these data with the value of the services derived from them.

Here we describe experiments where methods from Mechanism Design were used to elicit a truthful valuation from users for their Internet data and for services to screen people for medical conditions. In these experiments, 880 people from around the world were asked to participate in an auction to provide their data for uses differing in their contribution to the participant, to society, and in the disease they addressed. Some users were offered monetary compensation for their participation, while others were asked to pay to participate. 

Our findings show that 99\% of people were willing to contribute their data in exchange for monetary compensation and an analysis of their data, while 53\% were willing to pay to have their data analyzed. The average perceived value users assigned to their data was estimated at US\$49. Their value to screen them for a specific cancer was US\$22 while the value of this service offered to the general public was US\$22. Participants requested higher compensation when notified that their data would be used to analyze a more severe condition. They were willing to pay more to have their data analyzed when the condition was more severe, when they had higher education or if they had recently experienced a serious medical condition.

Our findings show that it is possible to place a monetary value on health-related uses of highly personal data. Such uses are valued by users and their value is approximately half that of their data. Our methodology can be extended to other areas where sensitive data may be exchanged for services to individuals and to society, while our results suggest that future services utilizing individual's Internet data could be viable.
\end{abstract}

\input{intro.tex}
\input{methods.tex}
\input{results.tex}
\input{discussion.tex}

\bibliographystyle{aaai}

\input{references.bbl}
\input{appendix.tex}

\end{document}

%% file: intro.tex
\section{Introduction}

\noindent Data is a valuable asset for organizations in a data-driven economy \cite{dewar2017}, but data holders can abuse this asset, one result of which is the possible breach of an individual's privacy. In response to the latter and to other issues stemming from amassing data by companies, the European Union's General Data Protection Regulation (GDPR) which came into effect in May 2018 gives control to individuals over their personal data (\url{eugdpr.org}), thus attempting to shift the balance between the value of data for individuals and for enterprises.

Though it is evident that data has value, quantifying it is difficult. In the past, researchers have attempted to put a monetary value that individuals assign to their photos \cite{bigmac}, browsing \cite{bigmac}, data from home appliances \cite{Kugler2018} and location \cite{locationprivacy}. Additionally, researchers have proposed “active choice” models which offer users the option of payment through monetary transactions or through disclosure of personal information \cite{Pricingprivacy}. 

One area where the value of the data has not been measured is healthcare. This is a major lacunae because healthcare is, arguably, the area in which data has the highest value to individuals. One the one hand, the high value is due to the potential for damage if privacy is breached, and on the other hand, for companies and individuals, because of the potential uses for creating new screening, treatment, and insights from data.

In recent years data from online services has proven a boon for medical research. For example, the website PatientsLikeMe (\url{www.patientslikeme.com}) lets people who are suffering from one of a number of diseases connect with others that suffer from similar problems for social support and for advice. The data posted by users of this site has enabled researchers to test the efficacy of treatments which would have been difficult to test through other methods, for example in rare diseases such as ALS \cite{wicks2011}. 

Another source of data from online services are people's queries on Internet search engines. These queries and the interactions of users with these search engines have been used to screen for severe medical conditions such as Parkinson's disease \cite{allerhand2018,white2018}, ovarian and cervical cancer \cite{soldaini2017}, lung cancer \cite{white2017}, pancreatic cancer \cite{paparrizos2016} and diabetes \cite{hochberg2019}. Similarly, social media postings have been used to predict depression \cite{dechoudhury2013} and diagnose autism \cite{ben2016}. 

Thus, people's data in online services can be used to create new medical services, e.g., screening tests for disease, but these data could potentially compromise individual privacy. Therefore, it is important to assess the value people assign to these data, vis-a-vis the value they perceive to be gained from such novel services. 

We note that the value of these services need not be limited to direct monetary value to the individual. People are often willing to donate money, time, effort, data \cite{DataDonation} and even organs (e.g., blood and kidneys) to help others in need, presumably in exchange for societal benefit \cite{bloodDonation}, personal satisfaction \cite{bloodDonation}, or to improve their social standing \cite{bloodDonation}. Thus, measurement of the value of data needs to account for the many facets of its perceived value to people, and to offset this value with the potential harm which might be caused by mishandling thereof.

In this paper we attempt to measure the value people assign to their search queries on Google or Bing (henceforth search logs) for medical uses. 

Measuring the value of data is a difficult undertaking because it requires eliciting a truthful value from people. People who are asked to provide unverifiable information  may misreport their valuations if they are skewed towards privacy, without incurring any cost. Thus, here we applied Mechanism Design \cite{nisan2001algorithmic} approaches, utilizing two forms of auction, a reverse Vickrey auction in one population and a Vickrey auction in another. In the reverse Vickrey auction we treat search logs as goods for sale, and design an auction with multiple sellers (agents) and one buyer (the authors of the paper). The buyer informs sellers that he is interested in at most $X$ units of the goods, and is willing to pay at most $r$ dollars for each unit; $r$ is known as the reserve price, and may be hidden. Each seller declares her bid (the required price for selling the goods), and the buyer buys $X$ units with the lowest bids, assuming the bids are below the reserve price and pays the minimum between $r$ and the $X+1$ lowest bid. 

Similarly, in the Vickrey auction we ask participants to provide their search logs and demand that they pay the experimenters to analyze them. In this case, the authors are sellers of an analysis service willing to process at most $X$ units of the goods (search logs), and are willing to accept the $X$ units offered at the highest price, if it is over $r$, a minimal (possibly hidden) reserve price. The details of the experiment and its modeling are described in Section \ref{sec:methods}.

The class of these mechanisms is \textit{dominant strategy incentive compatible} \cite{Vickrey}, meaning that it is in a participant's best interest to bid her honest value for the goods, regardless of $X,r$ or the bids declared by the other sellers. In our study, by letting participants  believe that we are willing to pay for their information, we are guaranteed, theoretically, that the elicited values are truthful. Importantly, we debriefed the participants after the experiment, and that too led to interesting findings (see Section \ref{sec:discussion}).  

Our paper provides several contributions. First, we develop a novel method for measuring and modeling the monetary value of data and data services. Second, we show that people assign a high value to health data and to health-related services. The value of the latter is such that almost half the people are willing to pay for these services, even when these services are provided to the public, without direct benefit to themselves. Our models suggest that the perceived value of data is approximately twice as that of the proposed health service using these data.

%% file: methods.tex
\section{Methods}\label{sec:methods}

\subsection{Data collection}

We conducted an online assessment of the value people assign to their internet search history data for its use in medical purposes by running an auction / reverse auction,  extracting people's valuations for health services and data in 8 different conditions.
Participants were recruited to participate in an online questionnaire through two crowdsourcing platforms, Mechanical Turk and Prolific Academic. The two platforms differ in the demographics of the workers employed in each, their geographic reach, and the attention of workers to the task \cite{peer2017}. For participating in the study, the participants who completed the questionnaire received US\$1.50.

At the beginning of the questionnaire we provided participants with background information, stating that search logs have been used for medical purposes. 

We then requested participants to consent to participate in the study and to provide their birth year, gender, country of residence, level of education and yearly income. They were also required to indicate if they have recently suffered from a serious medical condition and/or are currently suffering from it. 

We informed participants of the {\bf stated goal} of the questionnaire, which was to screen participants for further study in which those chosen would provide their search history on Bing and Google so that experimenters could use it towards one of 4 goals: 

\begin{enumerate}
\item {\bf Benign medical condition, public good}: Measure and report the rate of flu virus in the participant's country.
\item {\bf Severe medical condition, public good}: Develop a model to detect thyroid cancer using their data and apply it to people in the participant's country.
\item {\bf Severe medical condition, personal good}: Apply a model for detection of thyroid cancer to the participant's data and report the result to them.
\item {\bf Severe medical condition, public and personal good}: Develop a model to detect thyroid cancer, apply it to the participant's data and report the result to them.
\end{enumerate}

Participants were randomized into {\bf one} of the 4 goals, as well as to one of two additional conditions: Either the participant was asked how much money (in US dollars) they would require the experimenters to pay for their search history to be used for the stated goal or how much money participants would pay the experimenters to use their search history for the stated goal.

Thus, each participant was randomized into one of 8 experimental conditions (4 goals, pay or be paid).

As noted above, our goal was to elicit truthful evaluations from participants. Therefore, to check the willingness to pay, in the experiment we used a Vickrey auction \cite{Vickrey} with hidden reserve price. In this auction the $k$ highest bidders who submitted a bid higher than a reserve price $r$, win and pay the maximum between the $k+1$ highest bid and $r$. In our case, $r$ is hidden. 

For checking desire to be paid, in the experiment we used a reverse Vickrey auction with hidden reserve price. In this auction the $k$ lowest bidders who submitted a bid lower than a reserve price $r$, win and are paid the mimimum between the $k+1$ lowest bid and $r$. Both auctions are truthful, regardless of the reserve price. In both auctions we used reserve prices so that no winners will be selected; this is obtained by having an exorbitantly high reserve price (say \$10K) in the first condition and a very low reserve price (e.g., some negative number) in the second one.
While hidden reserve prices are common practice, we also included a description of the way they have been chosen in a debrief to the participants, which was well accepted, as described below.

Thus, participants were informed that the study, conducted by Anonymous Commercial Company (name removed for anonymous review),  was a first part of a two-staged study. In the first part, the questionnaire would be shown to 1000 people. Among those, the 100 people who requested the smallest amount of money for their data, if it were below a maximal threshold (or offered the largest amount if it were above a minimum threshold, in the second condition) would be contacted for the second stage of the study. 

Finally, because the study might be perceived as involving deception, we indicated to participants that we would be providing a debriefing on the study after they completed the questionnaire. Participants who indicated their interest in receiving the debriefing and get paid US\$0.10 for it were provided with the debriefing several weeks after the study was completed.

This study was approved by the Microsoft Institutional Review Board. See Appendix for the full questionnaire.

Participants who read the consent form but did not want to take part in the study and participants who did not complete the questionnaire were removed from the data.

\subsection{Modeling}

Let $V_{pu}^i$ and $V_{pr}^i$ be the valuation of  participant $i$ to the goods (public or personal, respectively), and let  $V_S^i=V_{pu}^i+V_{pr}^i$ be the total value for participant $i$ to the service. Let $V_D^i$ denote the cost for an agent in revealing his valuation (which may have privacy implications but also other ramifications). Let $P^i=V_S^i-V_D^i$ be the total valuation of participant $i$. Notice that $P^i$ is the value to be reported if/when he/she participates in a truth-revealing auction.  
Notice that under the above terminology  an equation of the form $P_i=V_S^i-V_D^i$ holds for each participant $i$; in the experiment in which participants are offered payments the reported bid $-P^i$ would be non-negative and in the experiment in which participants are asked to pay the reported bid  $P^i$ would be non-negative. The additive multi-attribute structure of utility we use is classical in economics and the characterization of conditions justifying it appears already in classical studies \cite{debreu1960}. 


Under the assumption that $V_{D}$, $V_{pr}$ and $V_{pu}$ are independent and that $P$ is a linear combination thereof, all proposed transactions (questionnaire responses) can be jointly represented using a linear model  where the dependent attribute is $P$ and the independent values are indicators of whether the service was offered in the transaction. Since all transactions involved data, $V_{D}$ is present in all transactions. 
Specifically, each transaction is in the form of $P^i=V^i_{pu}X^i_{pu}+ V^i_{pr}X^i_{pr}-V^i_{D}$, where $X^i_{pu}$ and $X^i_{pr}$ are indicators determined by the condition that user was shown.

The linear coefficients ($X^i_{pu}$ and $X^i_{pr}$) obtained as a solution to the equations above represent the average population valuation for the services and for the search log. More specifically, the obtained coefficients can be estimated through linear regression and the bias term will thus correspond to the average valuation for revelation of the search logs.

When modeling our data to understand demographic correlates therein, and to account for the skewed distribution of the requested and offered amounts we transformed the non-zero amounts using a log transform before modeling them. Participant's level of education was transformed into a continuous number based on the number of years needed to attain each level of education. For example, participants that indicated high-school as their higher education level received the number “12”. Income level was transformed into the average amount in the range, e.g.,  participants that indicated US\$15,000--US\$30,000 as their yearly income level received the number US\$22,500. 

%% file: results.tex
\section{Results}

We recruited 482 participants through Mechanical Turk and 398 through Prolific Academic. Participants were successfully randomized into one of eight experimental conditions (chi$^2$ test, $P=0.47$). 

The average reported age of participants was 35 (s.d.: 11) years. The reported gender of 46\% was female, 54\% male (Less than 1\% did not provide it). Participants reported an average of 15 years of education (s.d.: 2). Education was correlated with age (Spearman $\rho = 0.14$, $P=5\cdot10^{-5}$) and with income (Spearman $\rho = 0.27$, $P<10^{-10}$). 

Participants from Mechanical Turk were predominantly from the US (85\%) and India (12\%), whereas those from Prolific Academic were recruited from 23 countries, the most common being US (28\%), UK (23\%), and Poland (9\%).

\begin{figure}[tb]
\centering
\includegraphics[width=7.5cm]{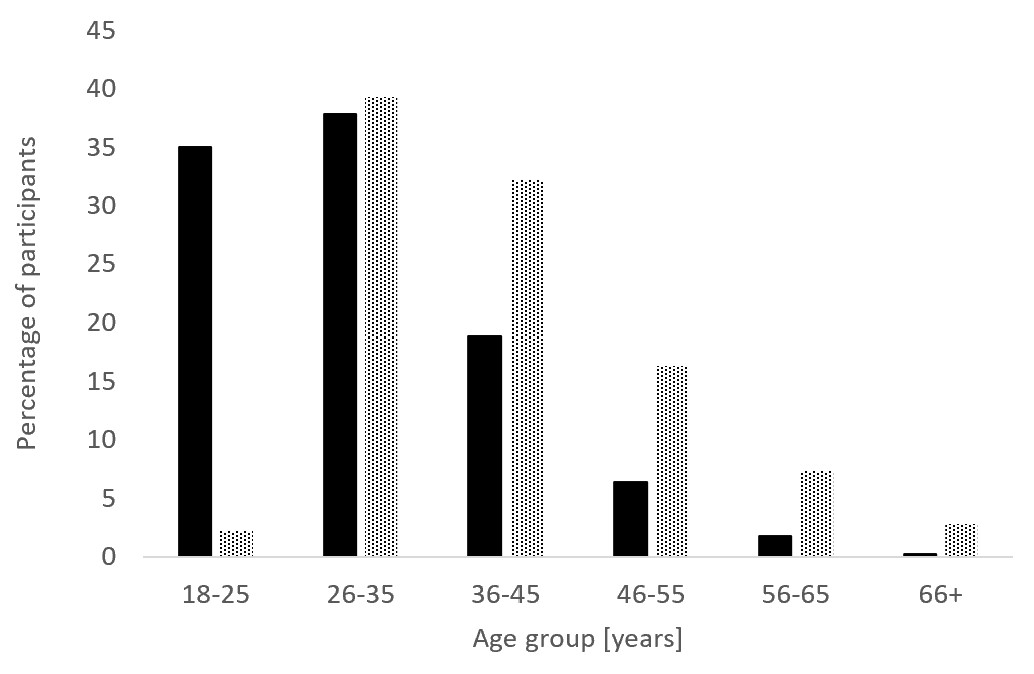}
\caption{Comparison of the ages of participants, by recruitment platform. Prolific Academic users are shown in black bars, and those from Mechanical Turk in gray.}
\label{fig:comparison1}
\end{figure}

\begin{figure}[tb]
\centering
\includegraphics[width=8.5cm]{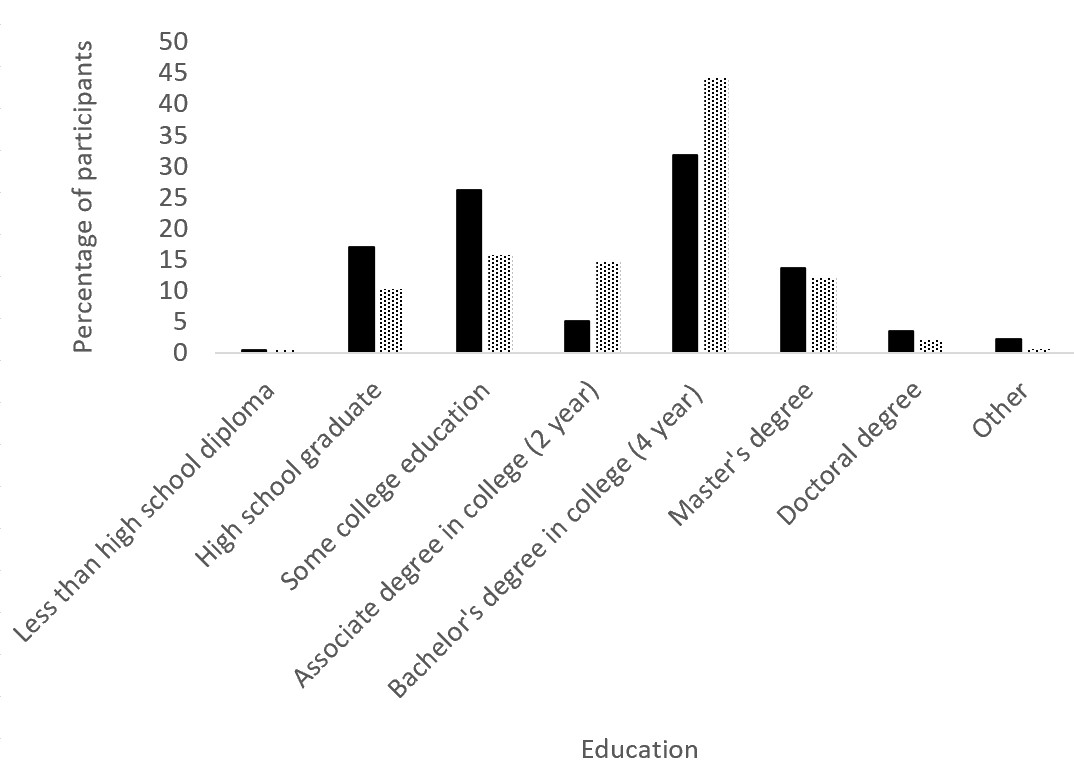}
\caption{Comparison of the education level of participants, by recruitment platform. Prolific Academic users are shown in black bars, and those from Mechanical Turk in gray.}
\label{fig:comparison2}
\end{figure}

\begin{figure}[tb]
\centering
\includegraphics[width=8.5cm]{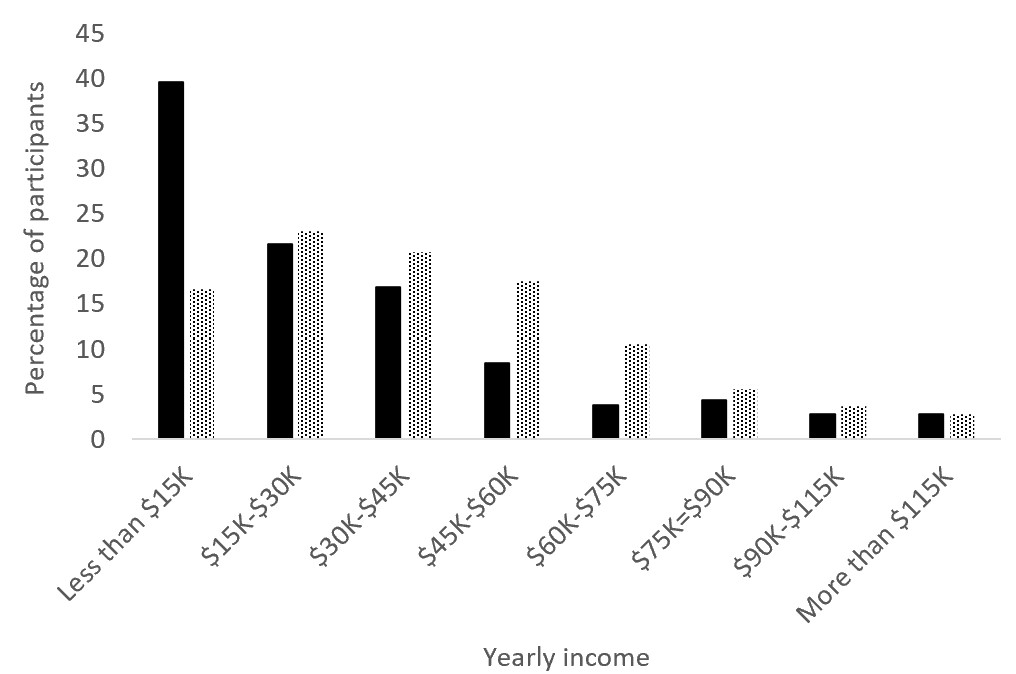}
\caption{Comparison of reported income, by recruitment platform. Prolific Academic users are shown in black bars, and those from Mechanical Turk in gray.}
\label{fig:comparison3}
\end{figure}

Figures \ref{fig:comparison1}, \ref{fig:comparison2} and \ref{fig:comparison3} compare the distribution of age, education, and income of participants, respectively, stratified by whether the participants were recruited from Mechanical Turk or from Prolific Academic. As the figures show, Prolific Academic participants were typically younger, with fewer years of  education and more were in the lower income bracket. 

\begin{figure}[tb]
\centering
\includegraphics[width=5.5cm]{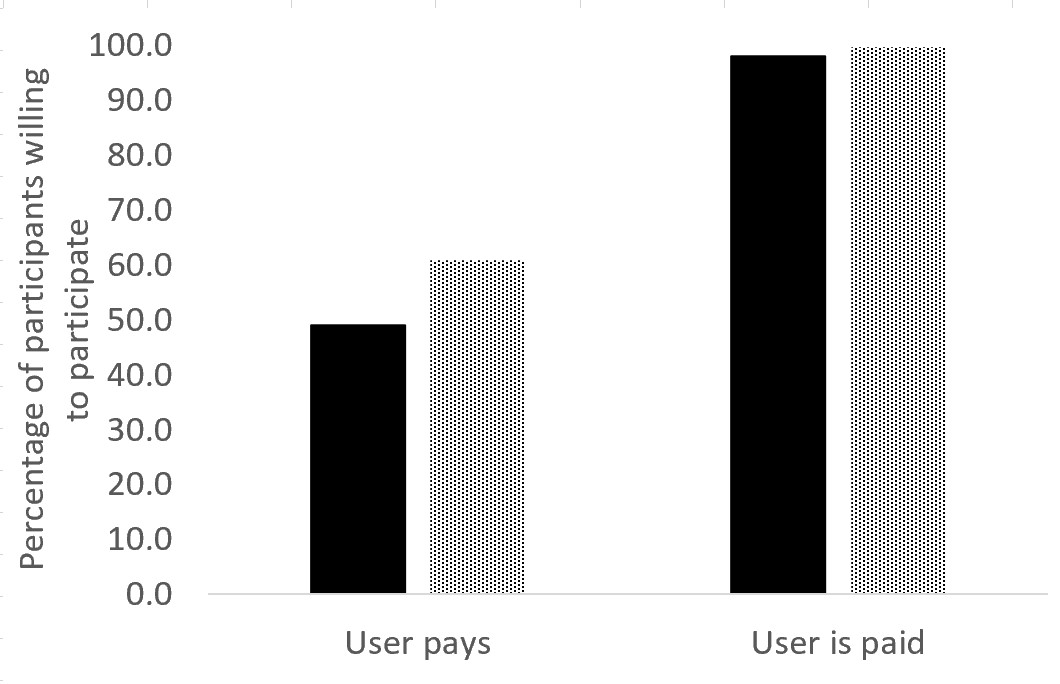}
\caption{Willingness to provide a monetary value, defined as a willingness to bid a non-zero value. Prolific Academic users are shown in black bars, and those from Mechanical Turk in gray.}
\label{fig:participation}
\end{figure}

We excluded 20 participants who responded in under 10 seconds to the monetary value question and another 41 participants who offered or requested more than US\$1500 (9 offered a higher amount and 32 requested a higher amount). Thus, 819 responses (93\%) were analyzed. Among 419 participants who were asked to pay, 224 (53\%) were willing to pay more than US\$0.10. Similarly, among 400 participants who were offered money, 395 (99\%) asked for more than US\$0.10. There were only small differences in willingness to offer a non-zero value between participants from Mechanical Turk and Prolific Academic, as shown in Figure \ref{fig:participation}.

The 224 users who offered a payment less than US\$0.10 and the 5 users who requested less than US\$0.10 should be considered as {\bf censored users}. Censoring \cite{oxforddictionary} is the failure to observe a variable totally, its value being replaced by a lower limit (right censoring), or an upper limit (left censoring). In our case, a user who offered to pay zero dollars might have been willing to provide their data had she been offered payment, while a user to whom we offered payment and requested a zero amount might have been willing to pay money for their data. However, since we only asked each participant whether they were willing to pay or be paid, our measurements are censored. 

\begin{figure}[tb]
\centering
\includegraphics[width=8.5cm]{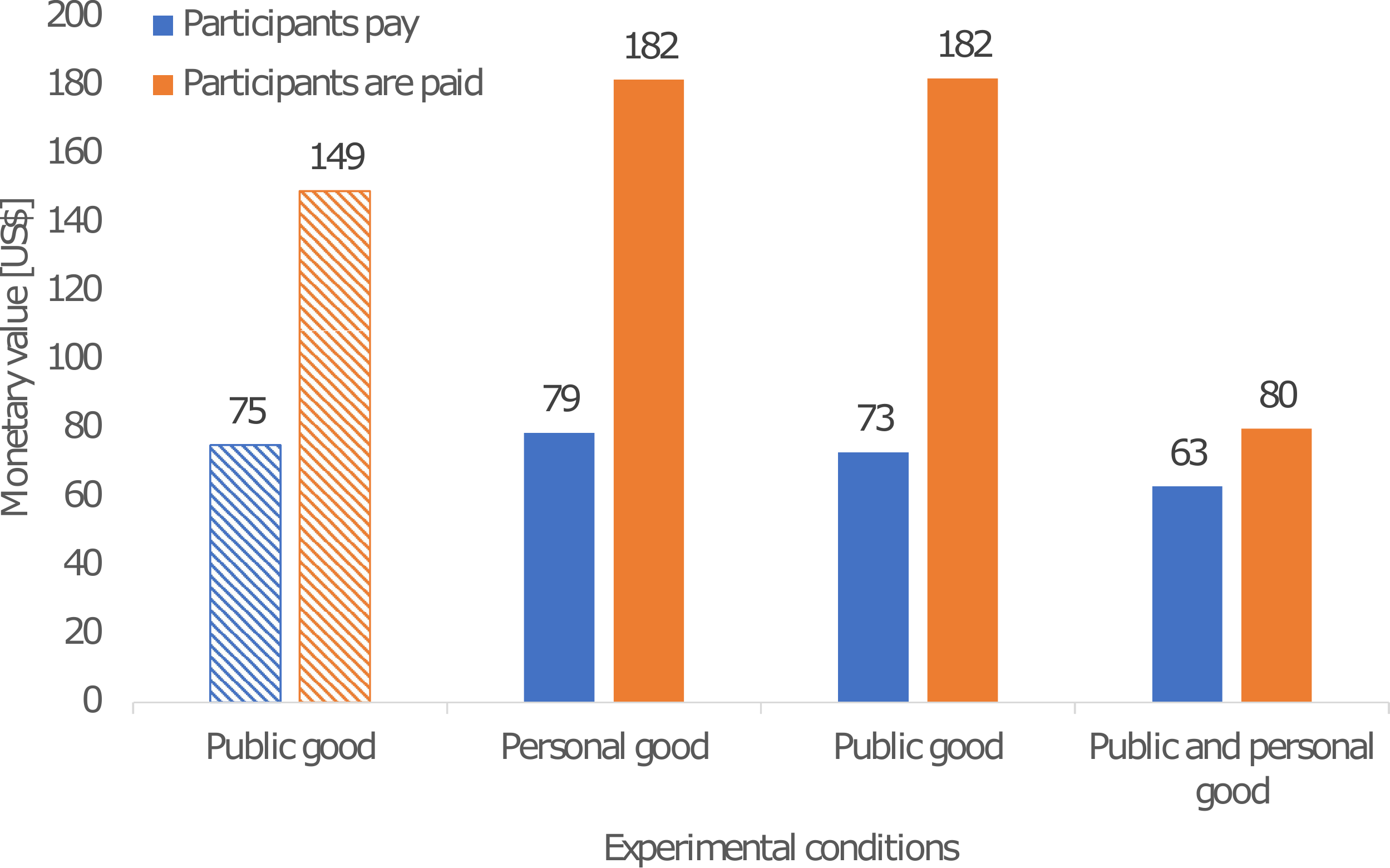}
\caption{Average bid values after removing censored users. Orange bars represent cases where participants were offered payment and blue bars cases where they were asked to pay. The left two bars are for the benign medical condition, while the other bars refer to the severe medical condition.}
\label{fig:ave_bids}
\end{figure}

A logistic regression model did not find any of the demographic variables statistically significantly associated with being censored. Henceforth we removed the censored users in our analysis. The average value users offered to give for analysis of their data was US\$38 (s.d.: 150) and the average value requested was US\$148 (s.d.: 269). Figure \ref{fig:ave_bids} shows the average bid values per question. As the figure shows, people request a larger amount than they are willing to pay for the service. Additionally, a larger amount is requested for the more severe conditions, but is not offered for such conditions. Strikingly, though the monetary value of public versus personal good is similar, when both are proposed people requested significantly less money and offered slightly less money. 


As described in Section \ref{sec:methods}, it is possible to estimate the average population valuation of the data and the two services by solving a linear regression problem. Thus, we modeled the untransformed value of $P$ using robust linear regression with 1\% of the largest outliers removed. The model is shown in Table \ref{tbl:robust}. As the table shows, the monetary value people attribute to their data is approximately \$49. Personal and public good are valued at approximately \$21. The severity of a condition is not statistically significantly correlated with the value offered by participants. 

\begin{table}[bt]
\centering
\begin{tabular}{lcc}
    \hline
    \hline
{\bf Variable} & {\bf Coefficients (SE)}    & {\bf P-value} \\
    \hline
Value of data   & 48.8 (12.4)   & 0.0002 \\
Severity        & -13.9 (10.1)  & 0.174 \\
Personal        & 22.1 (10.1)   & 0.028 \\
Public          & 20.4 (10.1)   & 0.042 \\
    \hline
    \hline

\end{tabular}
\caption{Model coefficients for predicting monetary value. Only non-censored observations are used (n=619). Model fit is $R^2=0.25$. }
\label{tbl:robust}
\end{table}

We modeled the monetary value of information, separately for participants who offered payment and those who were offered it, excluding censored users. For each group we constructed one model using only the experimental conditions (severity and type of good) and another which uses both these variables and demographic characteristics reported by users.

Table \ref{tbl:models} shows the model coefficients. As the table shows, users were willing to pay more for the use of their data if they had a higher education and if they had experienced a severe medical condition recently. People requested a higher compensation for their data if it were to be used to analyze the more severe condition, and reduced their demand if it contributed to both personal and public good. 

Note that in the model we included the severity as a variable, since it was  part of the transaction. However, its $p-value$ was not statistically significant, indicating that it had a negligible effect on the average population valuation.  
 
\begin{table*}[tb]
\centering
\begin{scriptsize}
    \begin{tabular}{l|cccc|cccc}
    \hline
    \hline
{\bf Variable} & \multicolumn{4}{c|}{{\bf User Pays}} & \multicolumn{4}{c}{{\bf User is Paid }} \\
& Coefficients (SE) & P-value & Coefficients (SE) & P-value & Coefficients (SE) & P-value & Coefficients (SE) & P-value\\
Severity & 0.149 (0.171) & 0.386 & 0.182 (0.168) & 0.281 & 0.272 (0.102) & 0.008 & 0.251 (0.102) & 0.014\\
Personal & -0.063 (0.161) & 0.696 & -0.090 (0.160) & 0.576 & -0.274 (0.104) & 0.009 & -0.246 (0.104) & 0.019\\
Public & -0.101 (0.165) & 0.542 & -0.131 (0.165) & 0.430 & -0.247 (0.102) & 0.016 & -0.236 (0.102) & 0.021\\
Medical status & & & 0.365 (0.133) & 0.007 & & & 0.009 (0.081) & 0.914\\
Education & & & 0.066 (0.027) & 0.017 & & & 0.004 (0.015) & 0.811\\
Income &  & & 0.000 (0.000) & 0.675 & & & 0.000 (0.000)  & 0.054\\
Age & & & -0.008 (0.006) & 0.150 & & & 0.006 (0.003) & 0.105\\
Gender & & & -0.020 (0.124) & 0.870 & & & 0.033 (0.075) & 0.659\\
    \hline
    \hline

\end{tabular}
\end{scriptsize}
\caption{Model coefficients for predicting monetary value. Two models are shown per condition: One where only the experimental conditions are used and the other where also user characteristics are included. Only non-censored observations are used.}
\label{tbl:models}
\end{table*}

As noted in Section \ref{sec:methods}, we offered participants a (paid) debriefing 4 weeks after they completed the questionnaire. Of 450 people who participated through Mechanical Turk, 363 (81\%) asked to receive the debrief and 353 of 369 (96\%) Prolific participants asked for it. Because of the way that the debrief was provided we could measure how many Prolific participants read the debrief. We found that 326 (92\%) read it. 

A few participants emailed us after the debrief. Notably, one wrote that ``[the study] totally had me fooled!". This would suggest that our methodology of a two-staged study so as to elicit truthful responses, was effective. 

%% file: discussion.tex
\section{Discussion}\label{sec:discussion}

People increasingly turn to the Internet when they have a medical condition, to diagnose it, learn about their options, and meet other people experiencing similar conditions \cite{yomtov2016crowdsourced}. The data they create during this process is a valuable source for medical research and for future health services. 

Here we show that when offered a compensation, people demand a high price for utilization of their data for health-related services (\$149 to \$182, depending on condition). The price is higher for use in screening for a severe medical condition, and is equal whether the data is used for personal or public good. The latter can be explained by participants perceiving the value of the offered service (apply an algorithm for thyroid cancer detection to their data) as mostly useful for research, not for the individual. Interestingly, when both services are offered, people request less than half the price of each separately. 

More than half (53\%) of participants were willing to pay the experimenters, in addition to providing their data. Higher value was offered for a service examining the severe medical condition and to personal good (over public good). Strikingly, a lower payment was offered if both goods were satisfied. Higher education and a recent experience of a serious medical condition were associated with higher payment.

When jointly modeled through a linear model described in Section \ref{sec:methods}, the value of public and personal good ($V_{pu}$ and $V_{pr}$) were found to be similar, at approximately US\$21. In contrast, the perceived value of search logs was estimated at approximately US\$49. Thus, the value of search logs ($V_D$) for health uses (as outlined in the questionnaire) is significantly higher than the value of search logs for other uses, as estimated in a previous study (25 Euro, approximately US\$28) \cite{bigmac}. Interestingly, the value of the two services together is roughly equal to that of the valuation of the data. This means that it would be possible for a service provider to obtain the needed data to create her service for the cost of offering the public and the individual screening tests, with little or no monetary compensation to users. 




There are several limitations in the way that the goal of data use was mentioned. First, we examined willingness to pay or be paid for a single service. In a real-life application users might be offered screening for multiple diseases, which may increase the perceived value to participants. Additionally, we did not specify if our request for data was a one-time request or for ongoing access, nor the retrospective length of time that the data would be accessed. Future research will examine the effect of these variables on the valuation of the data and the offered services.

In many ways, the willingness to share your personal data for personal and/or public use shares similarities with organ donation, in the sense that people might be more willing to share information with the possibility to help others if the default is sharing of such data. Past work has shown that the main difference among countries in the rates of organ donation can be explained  not by kindness nor altruism, but instead in the consent form and default option. In many countries, implementing an Opt Out Policy is the main reason of an increase in the rate of organ donations \cite{Defaults}. Similarly, we hypothesize that if by default personal data will be used for uses such as the ones outlined in this work, and only if the user chooses to ``opt out" will it be removed; more often than not users will allow their information to be analyzed, potentially helping the greater good.

Moreover, the likelihood of one choosing to be a living liver donor increases significantly if a personal acquaintance is in need of a liver or was in need of it \cite{liverDoner}. This is similar to our results which indicate that people who suffered from a medical condition themselves were willing to pay more for the use of their data than those who did not. 
However, care should be taken not to use data only from those who are biased towards data donation, as this could cause a bias in the models, both for the condition that they are suffering from and in other conditions. 

We believe that when people will be able to share their personal data for medical uses, one of the barriers that may prevent people from doing so will be the lack of supervision on the use of these data. This will be especially true if the default for use of data will be opt in. Thus, we envision an independent organization, similar to Institutional Review Boards, which will oversee proper use of data. Such an organization will weigh the use of personal information against the possibility of harm that can arise from its use; allowing personal information to be used (and possibly shared) when the greater cause outweighs the possible harm. 


\subsection{Acknowledgements}

Omer Ben-Porat and Moshe Tennenholtz are supported by the European Research Council (ERC) under the European Union's Horizon 2020 research and innovation programme (grant agreement number 740435).

%% file: appendix.tex
\section*{Appendix: Full questionnaire and debrief statement}

\subsection*{Questionnaire}

\textit{Demographic Questions:}

\begin{itemize}
\item What year were you born?
\item What is your gender?\\
	- Male\\
	- Female\\
	- Other\\
	- I prefer not to disclose
\item In which country do you currently reside?
\item What is the highest level of school you have completed or the highest degree you have received?\\
    - Less than high school diploma\\
	- High school graduate\\
	- Some college but no degree\\
	- Associate degree in college (2 year)\\
	- Bachelor's degree in college (4 year)\\
	- Master's degree\\
	- Doctoral degree\\
	- Other
\item What is your yearly income level?\\
    - Less than 15,000\$\\
    - 15,000\$-30,000\$\\
    - 30,000\$-45,000\$\\
    - 45,000\$-60,000\$\\
    - 60,000\$-75,000\$\\
    - 75,000\$-90,000\$\\
    - 90,000\$-115,000\$\\
    - More than 115,000\$
\item Have you suffered from a serious medical condition in the past year?\\
    - Yes\\
    - No
\item Are you currently suffering from a serious medical condition?\\
    - Yes\\
    - No
\end{itemize}

\textit{Experimental condition:}

(Each participant received only one question from the following):

\begin{itemize}
\item We want to measure and report the rate of flu virus in your country. 
\item We want to develop a way to detect thyroid cancer by analyzing your Bing or Google search queries and apply it to people in your country.
\item An algorithm will examine your data in order if it suggests that you have thyroid cancer and report the results to you.
\item We want to develop a way to detect thyroid cancer using your Bing or Google search queries. We will examine your data in order to see if it suggests that you have thyroid cancer and report the results to you.
\end{itemize}

(Each participant received only one question from the following):

\begin{itemize}
\item To do so, we would like to purchase your Bing or Google search queries in order to analyze them. We plan to recruit 1000 people to complete this questionnaire. Since we only require 100 responses, we will only be contacting the 100 people who requested the least amount of money (as long as it is less than our maximum).
For how much money (in US dollars) would you be willing to sell us your search queries for analysis? 
\item To do so, we would like to analyze your Bing or Google search queries.We plan to recruit 1000 people to complete this questionnaire. Since we only need 100 responses, we will contact the 100 people who were willing to pay the largest amount of money (as long as it is more than our minimum amount).
How much money (in US dollars) are you willing to pay so that we analyze your search queries?
\end{itemize}

\subsection*{Debrief statement}

Thank you for your participation in our study! Your participation is greatly appreciated.

As we noted in the description of our study, recent advances in medical research have shown that it is possible to diagnose serious medical conditions from the searches people make online through services such as Google and Bing. In the consent form of the study we informed you that the purpose of our study was to recruit a large group of people from whom we will collect their search history, which we will use to improve our ability to discover medical conditions.

In actuality, the goal of our study was to estimate the monetary value that users place on data which can be used to assist in medical research. The way in which the questionnaire was structured is known to elicit truthful responses from people, but the minimal prices we intended to offer were set up so that no people could be recruited for the second stage of the research.

Unfortunately, to properly test our hypothesis, we could not reveal these details to you at the time of the experiment.

We hope that the results of our study will help us offer these novel screening services to people around the world in the near future. We thank you for helping us in understanding the value people ascribe to these services.